\documentclass[%
reprint,
 amsmath,amssymb,
 aps,
]{revtex4-1}

\usepackage{dcolumn}
\usepackage{bm}

\usepackage{floatrow}

\usepackage[greek,english]{babel} 
\usepackage[T1]{fontenc}
\usepackage[utf8]{inputenc}
\usepackage{upgreek} 
\usepackage{lmodern} 
\usepackage{subcaption}

\usepackage{epsfig}
\usepackage{amssymb}
\usepackage{mathrsfs}
\usepackage{times}
\usepackage{color}
\usepackage{ifpdf}
\usepackage{graphics}

\usepackage[singlelinecheck=on]{subcaption}
\usepackage[textfont=normalfont,singlelinecheck=off,justification=raggedright]{caption}
\usepackage{soul}

\ifpdf
\usepackage{epstopdf}
\usepackage{url}
\fi
\usepackage{graphicx} 
\usepackage{amsmath}
\usepackage{amsfonts}
\usepackage{siunitx}

\usepackage[colorinlistoftodos,prependcaption,textsize=tiny,textwidth=1.7cm,color=green!25]{todonotes}

\newcounter{todocounter}
\newcommand{\bra}[1]{\ensuremath{\left\langle#1\right|}}
\newcommand{\ket}[1]{\ensuremath{\left|#1\right\rangle}}

\newcommand{\bibo}{BiB$_3$O$_6$ }

\begin{document}

\preprint{APS/123-QED}

\title{Quantum storage of single-photon and two-photon Fock states with an all-optical quantum memory}

\author{M.~Bouillard}
\affiliation{Laboratoire Charles Fabry, Institut d'Optique Graduate School, Universit\'e Paris-Saclay, 2 Avenue Augustin Fresnel, 91127, Palaiseau, France}

\author{G.~Boucher}
\affiliation{Laboratoire Charles Fabry, Institut d'Optique Graduate School, Universit\'e Paris-Saclay, 2 Avenue Augustin Fresnel, 91127, Palaiseau, France}

\author{J.~Ferrer~Ortas}
\affiliation{Laboratoire Charles Fabry, Institut d'Optique Graduate School, Universit\'e Paris-Saclay, 2 Avenue Augustin Fresnel, 91127, Palaiseau, France}

\author{B.~Pointard}
\affiliation{Laboratoire Charles Fabry, Institut d'Optique Graduate School, Universit\'e Paris-Saclay, 2 Avenue Augustin Fresnel, 91127, Palaiseau, France}

\author{R.~Tualle-Brouri}
\affiliation{Laboratoire Charles Fabry, Institut d'Optique Graduate School, Universit\'e Paris-Saclay, 2 Avenue Augustin Fresnel, 91127, Palaiseau, France}

\date{\today}

\begin{abstract}

{Quantum memories are a crucial element towards efficient quantum protocols. In the continuous variables domain, such memories need to have near unity efficiencies. Moreover, one needs to store complex quantum states exhibiting negative Wigner functions after storage. We report the implementation of an all optical quantum memory: the storage of single and two-photon Fock states with high fidelities has been realized. The Wigner functions of the reconstructed states shows negativity after $\sim\SI{0.75}{\micro\second}$ and $\sim\SI{0.61}{\micro\second}$ respectively for the single-photon and two-photon Fock states. This is, to our knowledge, the first demonstration of the storage of non-Gaussian states with more than one photon, representing a key step towards hybrid quantum protocols.}
 \end{abstract}

\maketitle

A main bottleneck towards large-scale quantum-optics protocols concerns the ability to provide simultaneously several resources issued from non-deterministic protocols. In this context, efficient quantum memories are an essential feature for the implementation of sophisticated operations. Over the last decade, numerous memories have been realized in order to store quantum states using vapor cells~\cite{ReimMichelbergerLeeEtAl2011}, cold atoms~\citep{BimbardBoddedaVitrantEtAl2014,ParigiDAmbrosioArnoldEtAl2015,Vernaz-Gris2018}, crystals~\cite{JobezLaplaneTimoneyEtAl2015,SaglamyurekJinVermaEtAl2015}... Those memories usually exhibit long storage time (in the \SI{}{\milli\second} scale), high fidelities but low efficiency, making them good candidates for discrete variables protocols, where quantum states are usually measured via correlation or anti-correlation between multiple detection events. When considering hybrid quantum protocols, with the use of continuous variables, the states are usually measured with homodyne or heterodyne detections. Unlike the discrete variables case, the measurement of field quadratures always leads to a result, regardless of the presence of photons in the observed optical mode: there is no “lost modes” in that case, and the efficiency has to be taken as one. Failed extractions have to be considered as vacuum, leading to a substantial drop of the fidelity. For that reason, most quantum memories developed so far are not suitable for continuous variables.

Continuous variables quantum protocols rely on non-Gaussian states containing more than one photon~\cite{RalphGilchristMilburnEtAl2003,LundRalphHaselgrove2008,LeeJeong2013}, and a main asset of optical cavity-based quantum memories is their potentiality to store such states with high fidelity~\cite{MakinoHashimotoYoshikawaEtAl2016}. A key parameter of the generation of non-Gaussian states is the negativity of their Wigner functions (defined as their minimal value), which can be used as a quantitative witness of non-classicality. Thus, the negativity of the stored states is an important parameter for characterizing quantum memories. In this paper, we experimentally demonstrate the quantum storage, in an optical cavity, of Fock states up to the two-photon level exhibiting negative Wigner functions after dozens of round trips. To our knowledge, this experiment represents the first storage of a non-Gaussian quantum state containing more that one photon, which is a crucial step towards quantum information protocols.

\section{An optical quantum memory for hybrid protocols}

\begin{figure*}
	\includegraphics[width=1\linewidth]{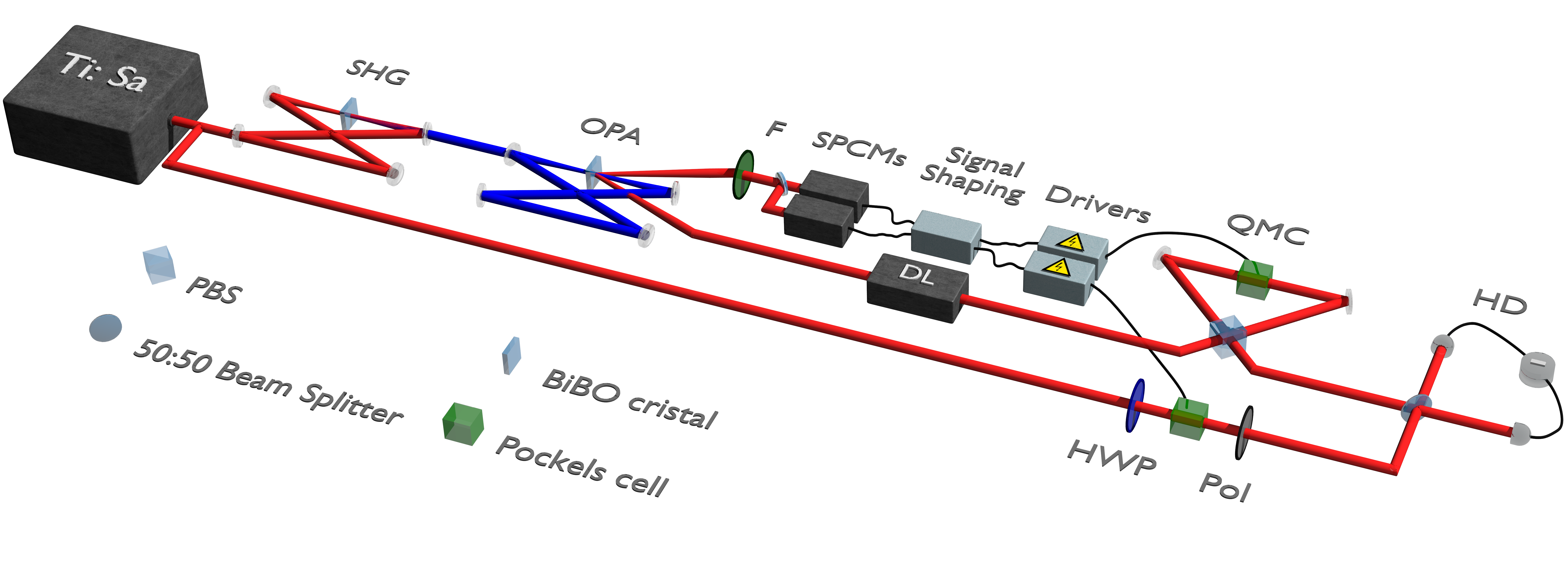}
    \caption{Scheme of the setup. Pairs of single-photon and two-photon states are produced by SPDC in a \bibo crystal placed in the OPA cavity where the pump beam is resonant. The heralding part of the photon pairs is spatially and spectrally filtered, and separated towards two SPCMs in order to announce the creation of the quantum states. The prepared state go through an optical delay line ($\sim\SI{200}{\nano\second}$) to account for the electronical delay of the Pockels Cells. Thanks to the Pockels cell, the quantum state is then stored inside the quantum memory for a given number of round trip, and sent to the homodyne detection for characterization. F: spectral and spatial filters, DL: delay line, PBS: polarizing beam-splitter, HWP: half-wave plate, Pol: Polarizer.}
    \label{scheme}
\end{figure*}

A scheme of the quantum memory is represented in figure~\ref{scheme}: it consists in a cavity with a thin film polarizing beam splitter (PBS), represented as a cube in the figure \ref{scheme}. A Pockels cell placed inside the cavity allows to switch the polarization of the light. Typically, a light pulse enters into the cavity with a horizontal polarization, the Pockels cell allows to rotate this polarization to vertical so that the pulse is then reflected by the PBS, and thus stored in the cavity. To extract the light, one just has to turn the polarization back to horizontal. The light can thus be delayed by an arbitrary number of round trips, with a synchronization that is naturally warranted, provided that the length of the cavity (\SI{3.9}{\meter}) matches the repetition rate of the laser (\SI{76}{\mega\hertz}).

 Recently such kind of protocols have been realized in order to synchronize single photons from two independent sources in the discrete regime~\cite{KanedaXuChapmanEtAl2017}. In this regime, an issue could appear when the polarization degree of freedom is used to store the light, as such cavities cannot store information encoded in polarization. In the continuous variables domain, the information can be encoded using the quadratures degree of freedom, and the polarization is no longer matter of interest for information encoding. However, with continuous variables, no measurement can be ruled out: the efficiency of a quantum memory thus has to be taken as unity, and the different losses will reduce the fidelity of the output states.

When using continuous variables, the non-Gaussian states are natural candidates for quantum protocols \cite{JeongKim2002,LloydBraunstein1999,RalphGilchristMilburnEtAl2003,LundRalphHaselgrove2008,LeeJeong2013,SilberhornRalphLuetkenhausEtAl2002}. In that direction, complex quantum states with coherent superpositions of Fock states with more than one photons are essential. For instance, Schrödinger cat states are very promising candidates for quantum computation and quantum information protocols, making them highly popular~\cite{OurjoumtsevTualle-BrouriLauratEtAl2006,ourjoumtsev2007generation,DelegliseDotsenkoSayrinEtAl2008,GerritsGlancyClementEtAl2010,etesse2015experimental,HuangJeannicRuaudelEtAl2015,SychevUlanovPushkinaEtAl2017}. Therefore, the ability to store states with more than one photon with a good fidelity is a crucial step.

Interestingly, this architecture for the quantum memory could allow not only to store quantum state but also to create quantum superposition such as NOON states or Schrödinger cat states. The Pockels cell in the cavity indeed offers the possibility to act as a quater-wave plate. In that case, the PBS in the cavity will act as a symmetric beam-splitter and, if two photons with orthogonal polarization are simultaneously present in the cavity, this will lead to the generation of a two-photon NOON state via photon coalescence. It has been demonstrated~\cite{EtesseKanseriTualle-Brouri2014,etesse2015experimental} that, from such a NOON state and with a homodyne measurement on one mode of the state, one can project the other mode in a Schrödinger cat state. This scheme could then allow to manipulate the state stored in the cavity, and therefore avoiding the use of multiple quantum memories and reducing the losses due to consecutive storages and extractions. The memory also allows to wait for the additional required resource, which is a single-photon in this example. The probability of success of such experiment is limited by the cavity losses as well as by the production rate of the single photons. If one fix $N$ as the maximum number of round trips allowed for the single-photon to keep a reasonable fidelity threshold, and if $p_{\ket{1}}$ is the probability of producing a single-photon Fock state by a source, the probability $P$ of synchronizing two single-photons with the cavity then reads:
%

\begin{equation}
P = 1-(1-p_{\ket{1}})^N
\label{eq:prob}
\end{equation}

\section{Experimental setup}

	\begin{figure*}
		\centering
		\begin{subfigure}{0.035\linewidth}
			\includegraphics[width=1.1\textwidth]{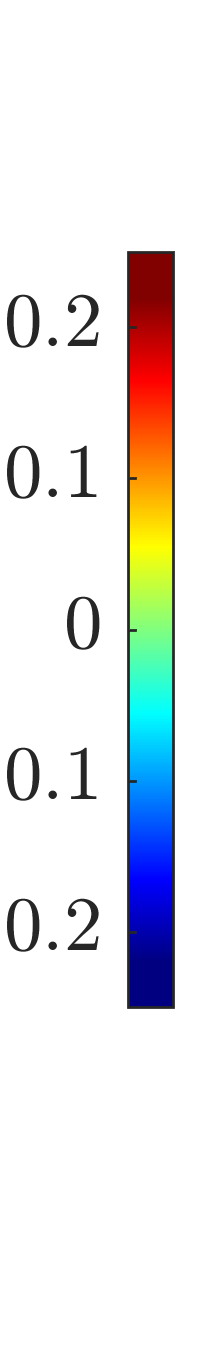}

			\includegraphics[width=1.1\textwidth]{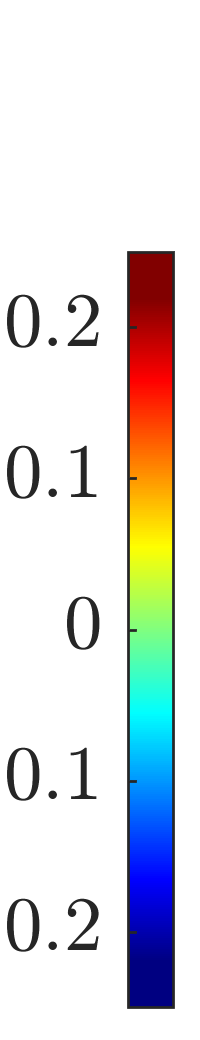}
            \vspace{5mm}
    	\end{subfigure}
        \begin{subfigure}{0.185\linewidth}
			\includegraphics[width=1.1\textwidth]{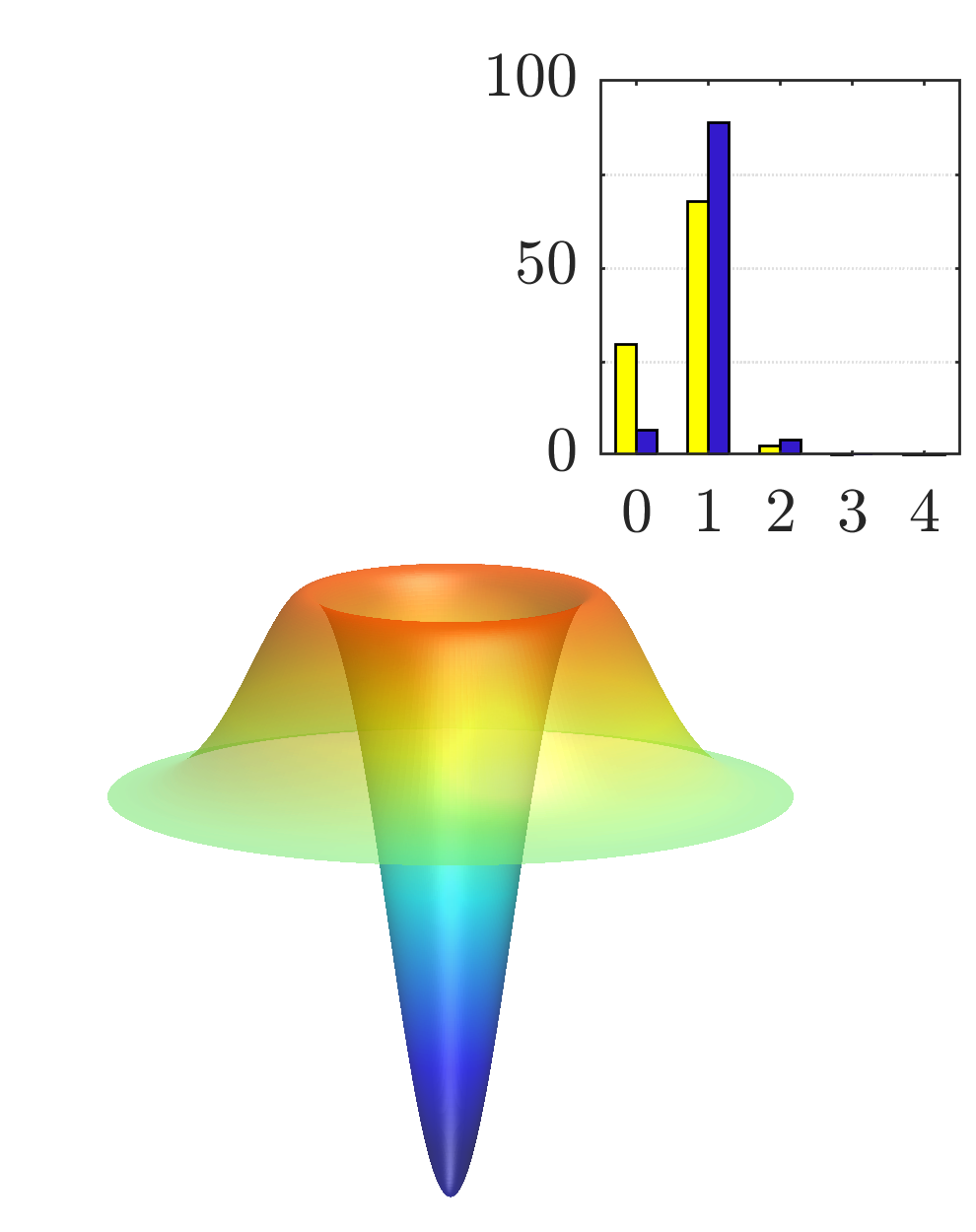}
            \centering
			\includegraphics[width=1.1\textwidth]{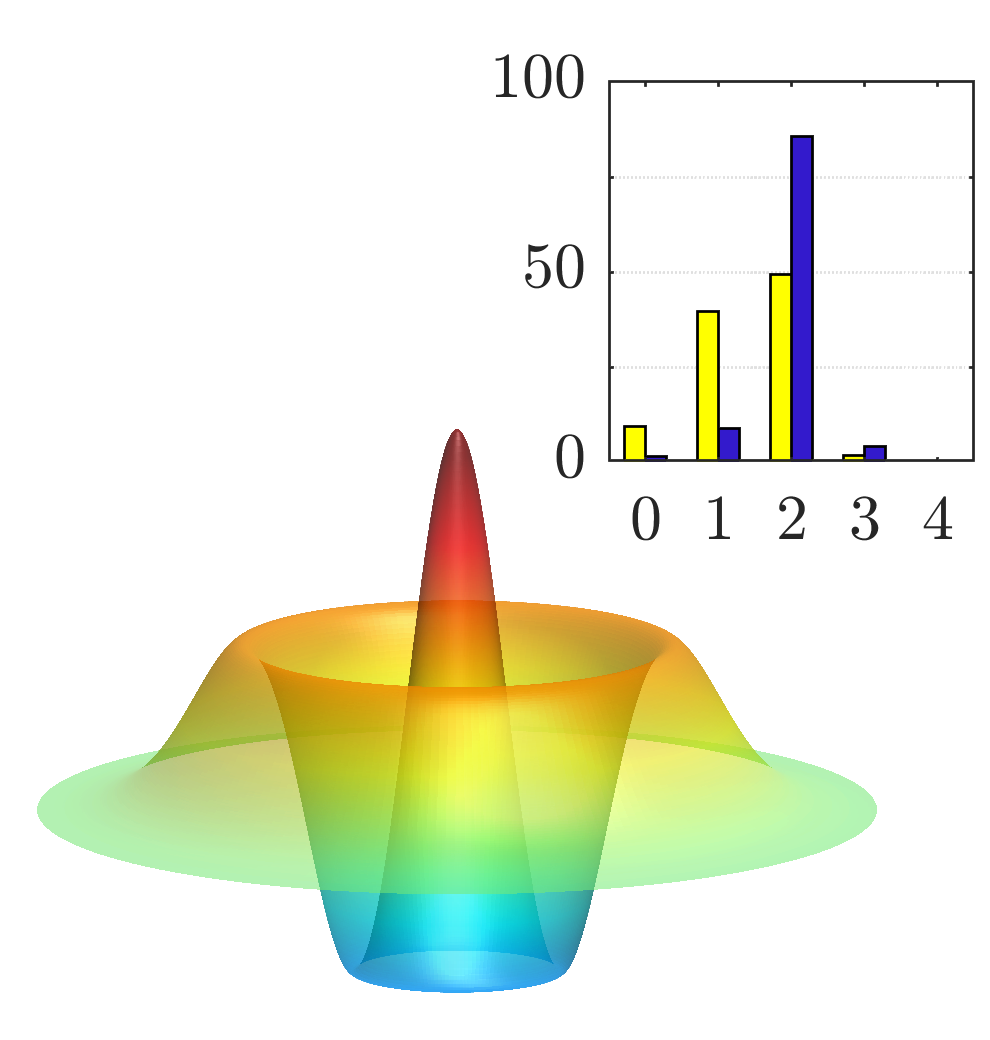}
    		\caption{0 round trips \\ (0 ns)}
    	\end{subfigure}
        \centering
		\begin{subfigure}{0.185\linewidth}
			\includegraphics[width=1.1\textwidth]{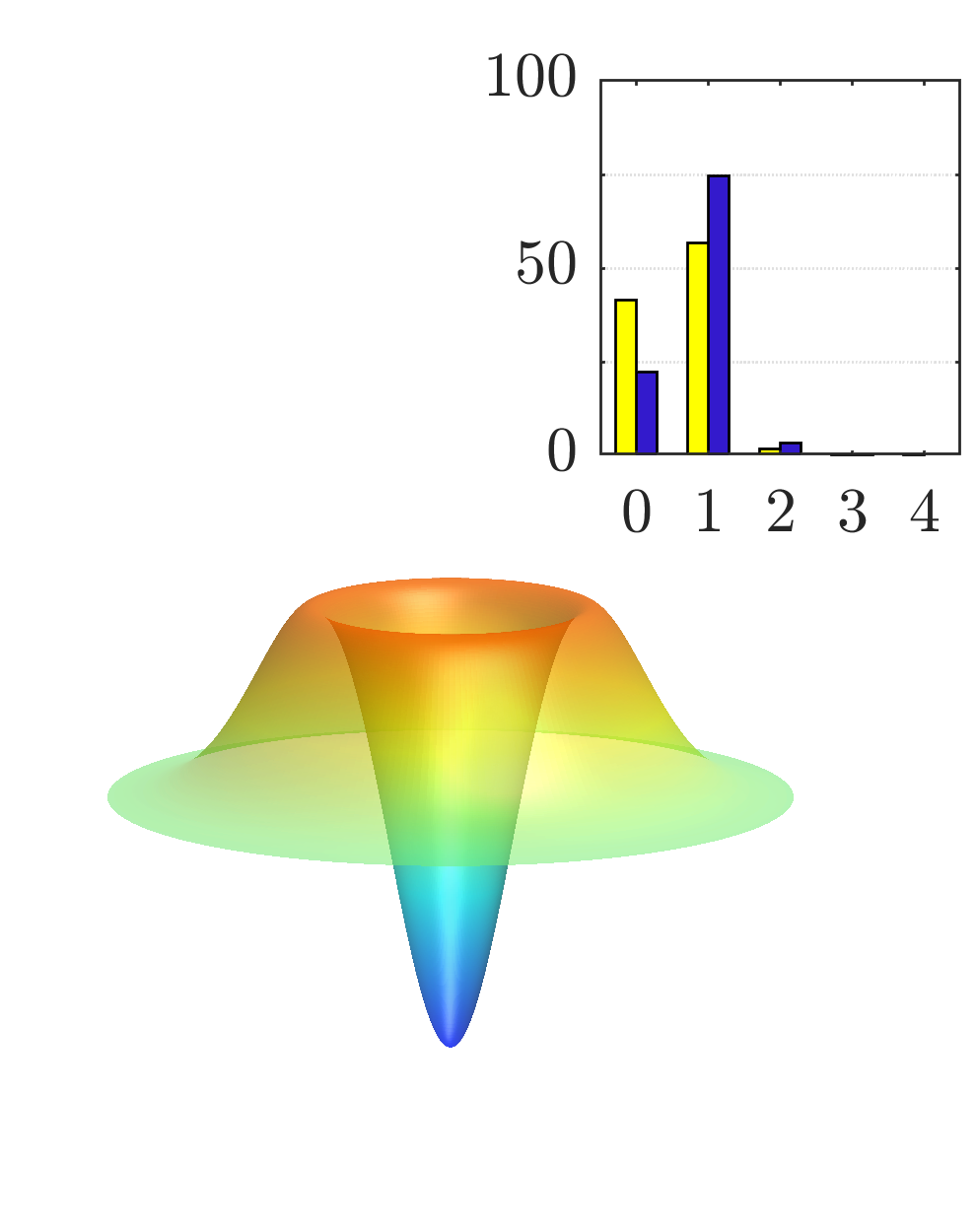}
			\includegraphics[width=1.1\textwidth]{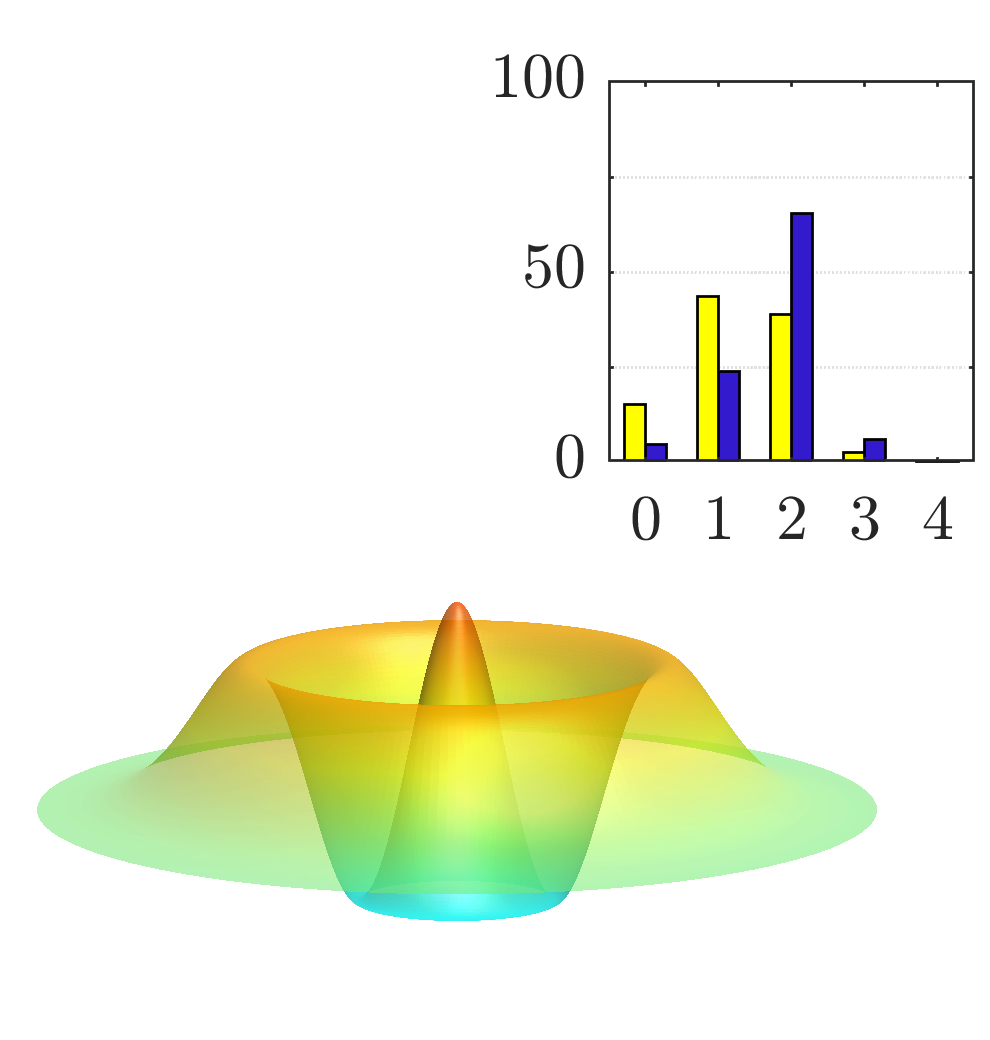}
    		\caption{10 round trips \\ (130 ns)}
    	\end{subfigure}
		\begin{subfigure}{0.185\linewidth}
			\includegraphics[width=1.1\textwidth]{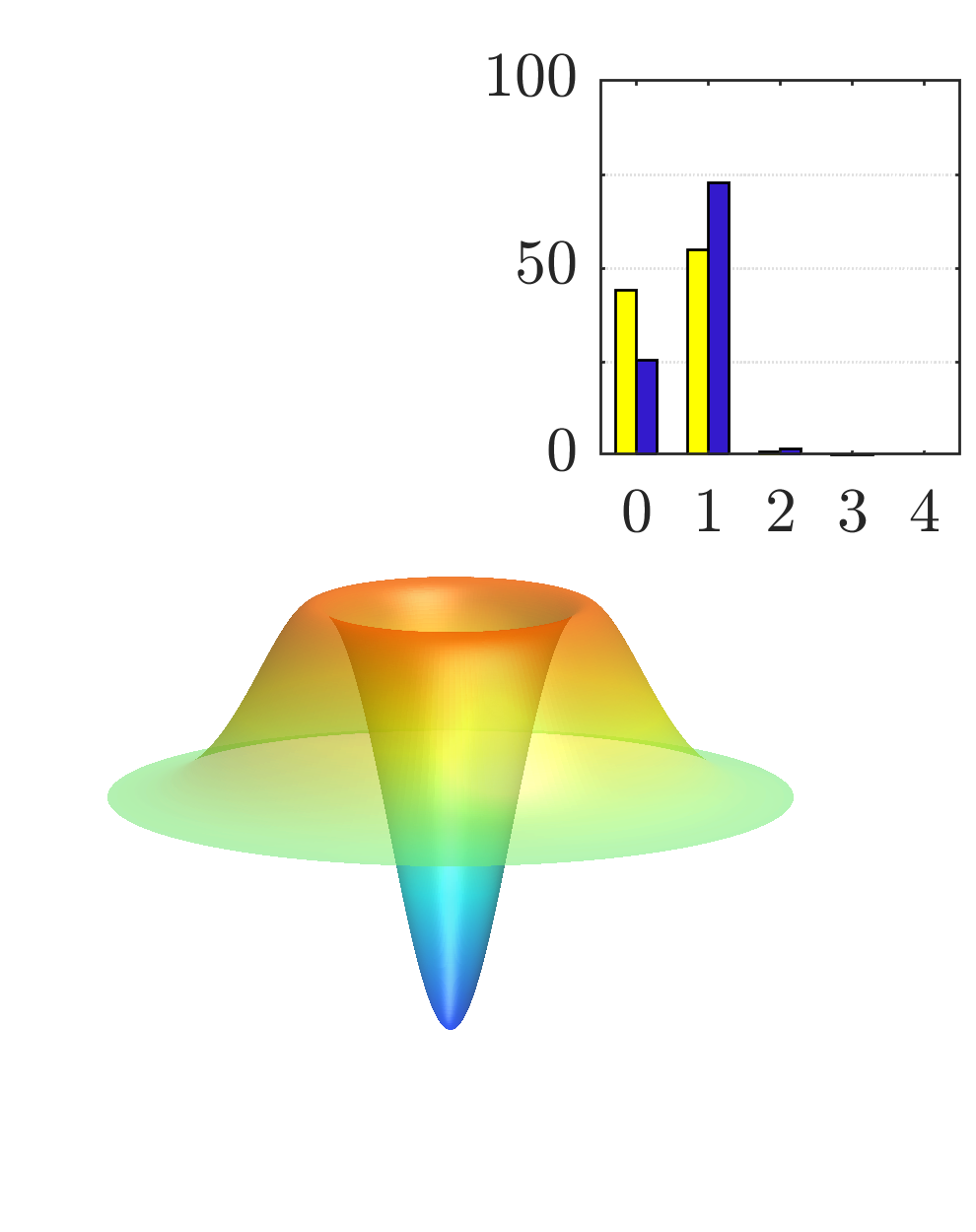}
			\includegraphics[width=1.1\textwidth]{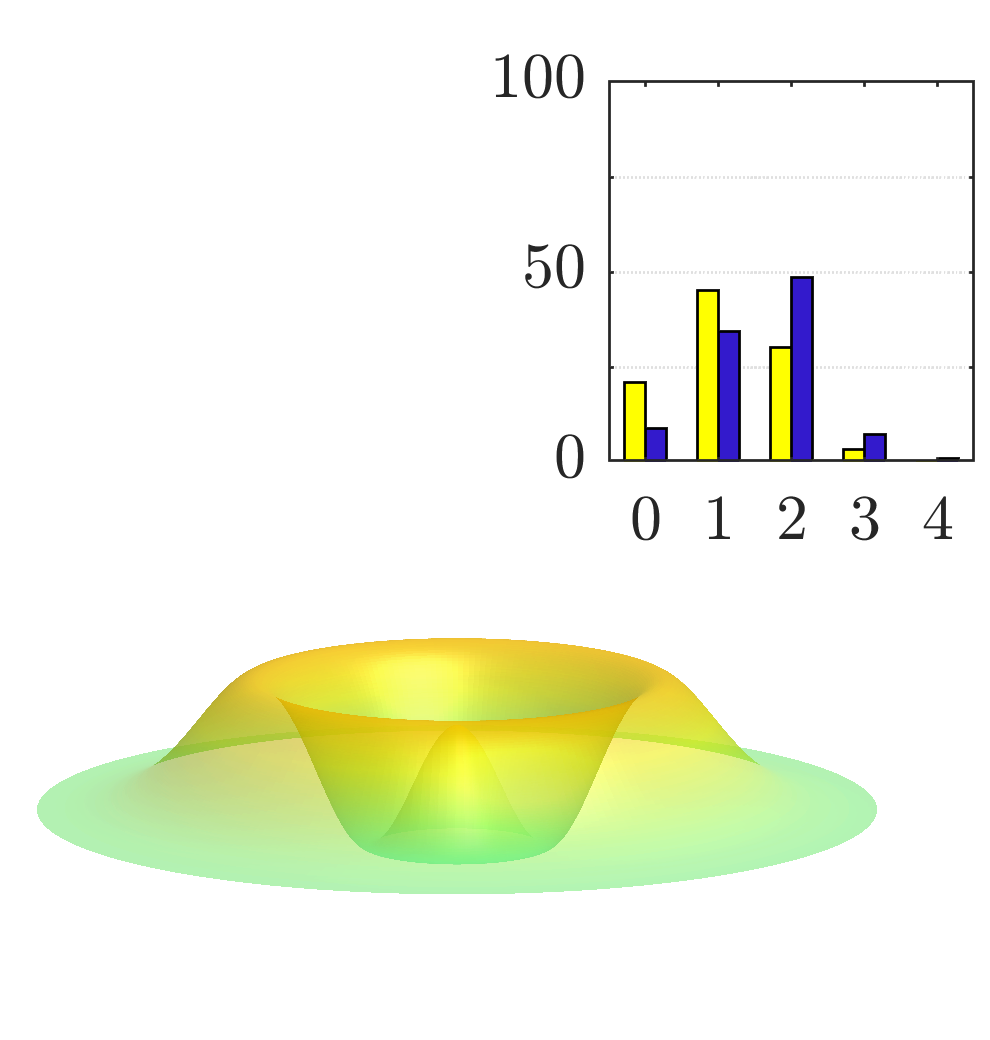}
    		\caption{20 round trips \\ (260 ns)}
    	\end{subfigure}
		\begin{subfigure}{0.185\linewidth}
			\includegraphics[width=1.1\textwidth]{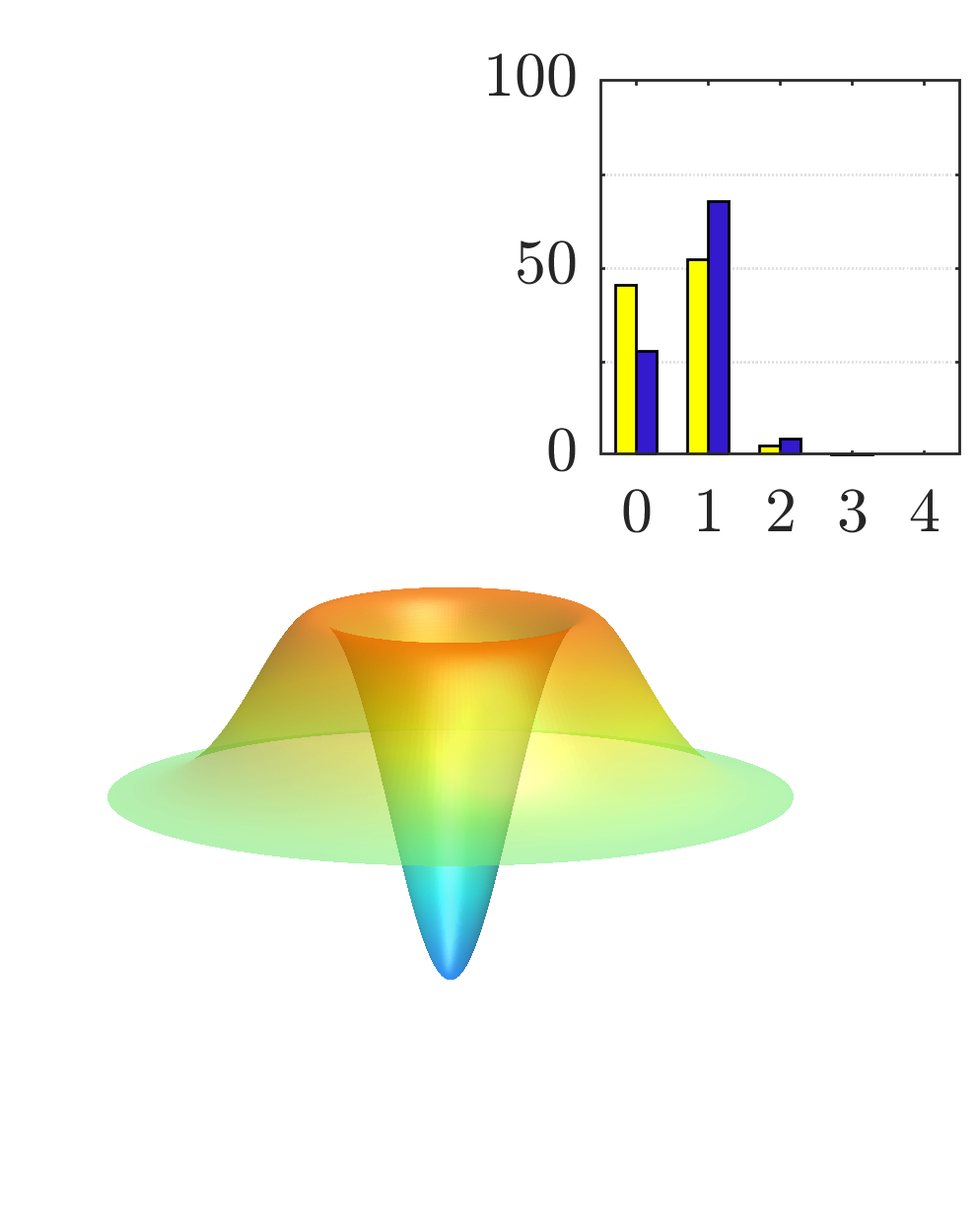}
			\includegraphics[width=1.1\textwidth]{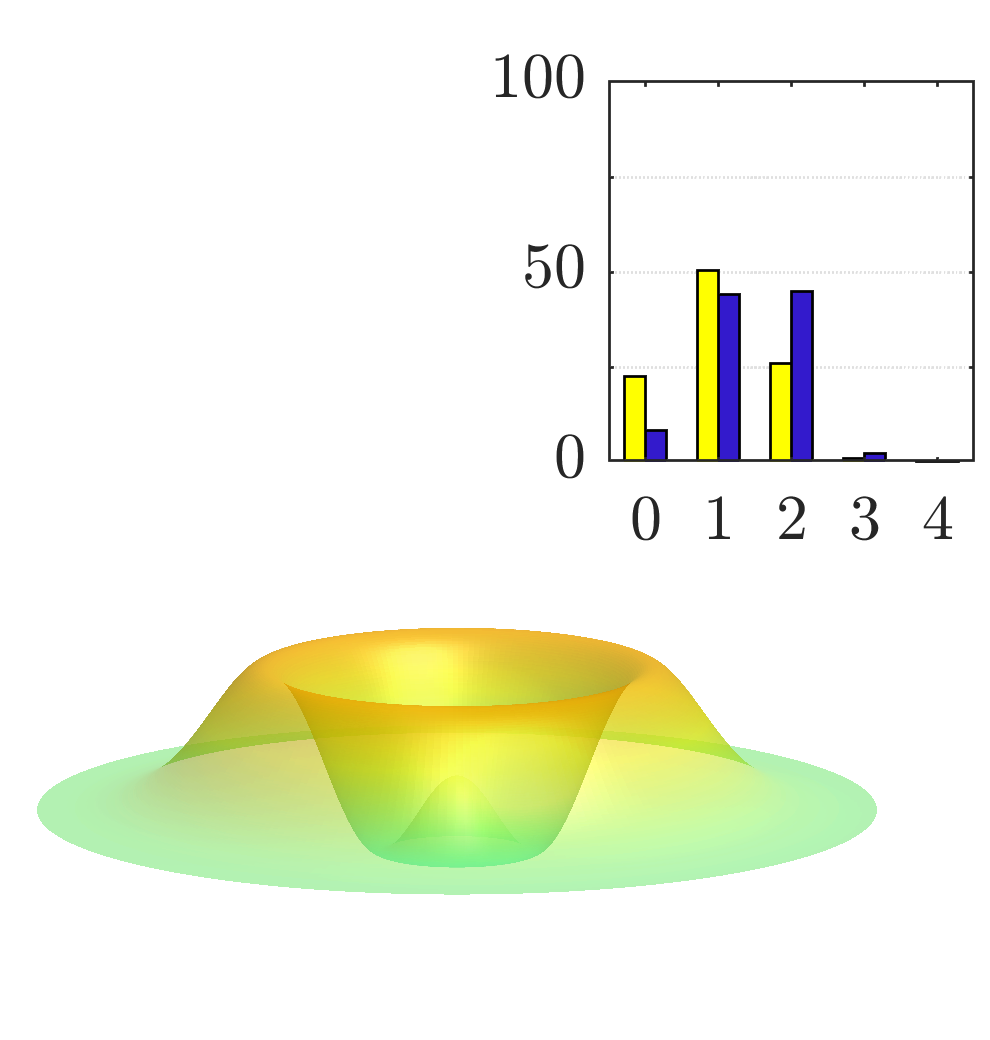}
    		\caption{30 round trips \\ (390 ns)}
    	\end{subfigure}
		\begin{subfigure}{0.185\linewidth}
			\includegraphics[width=1.1\textwidth]{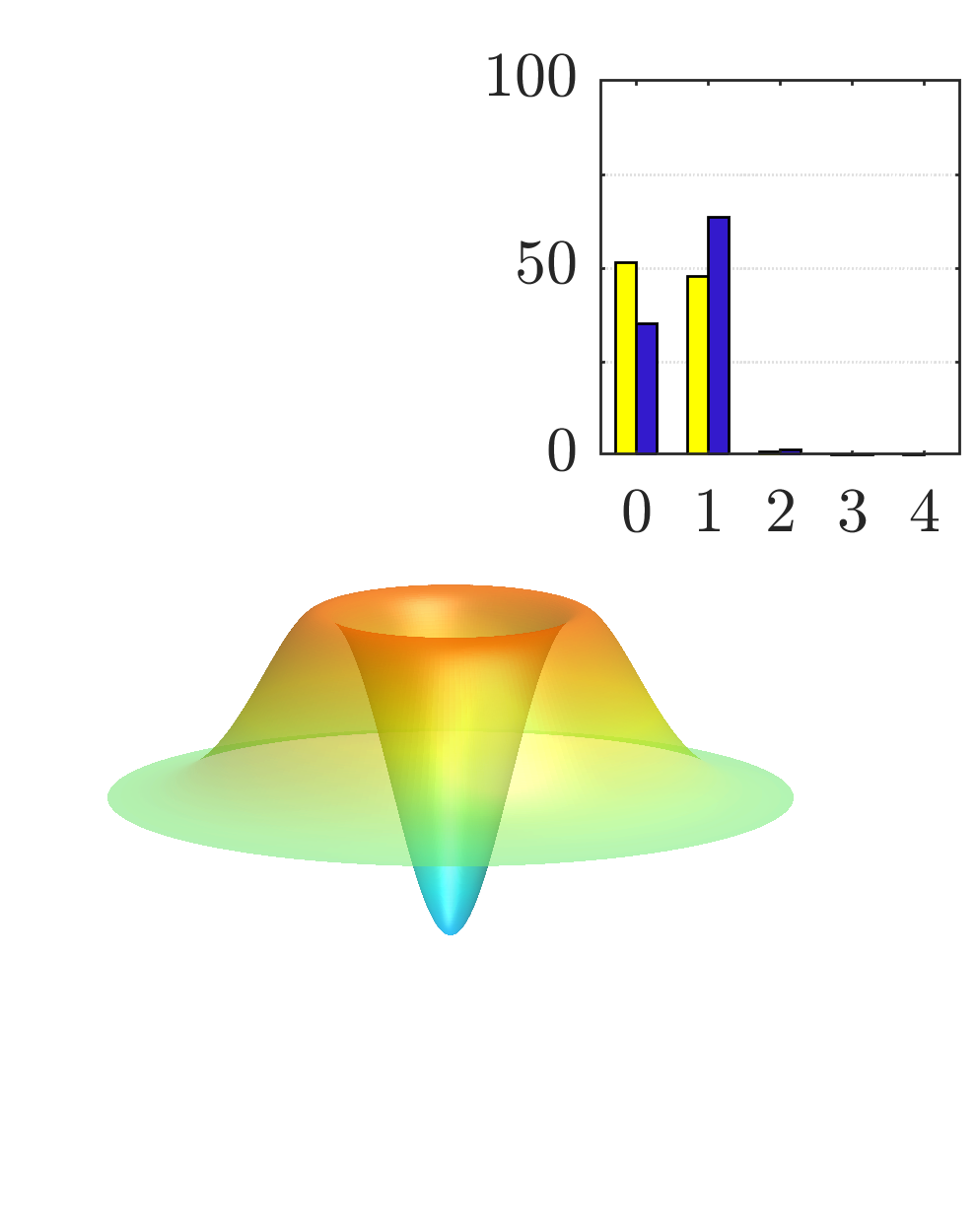}
			\includegraphics[width=1.1\textwidth]{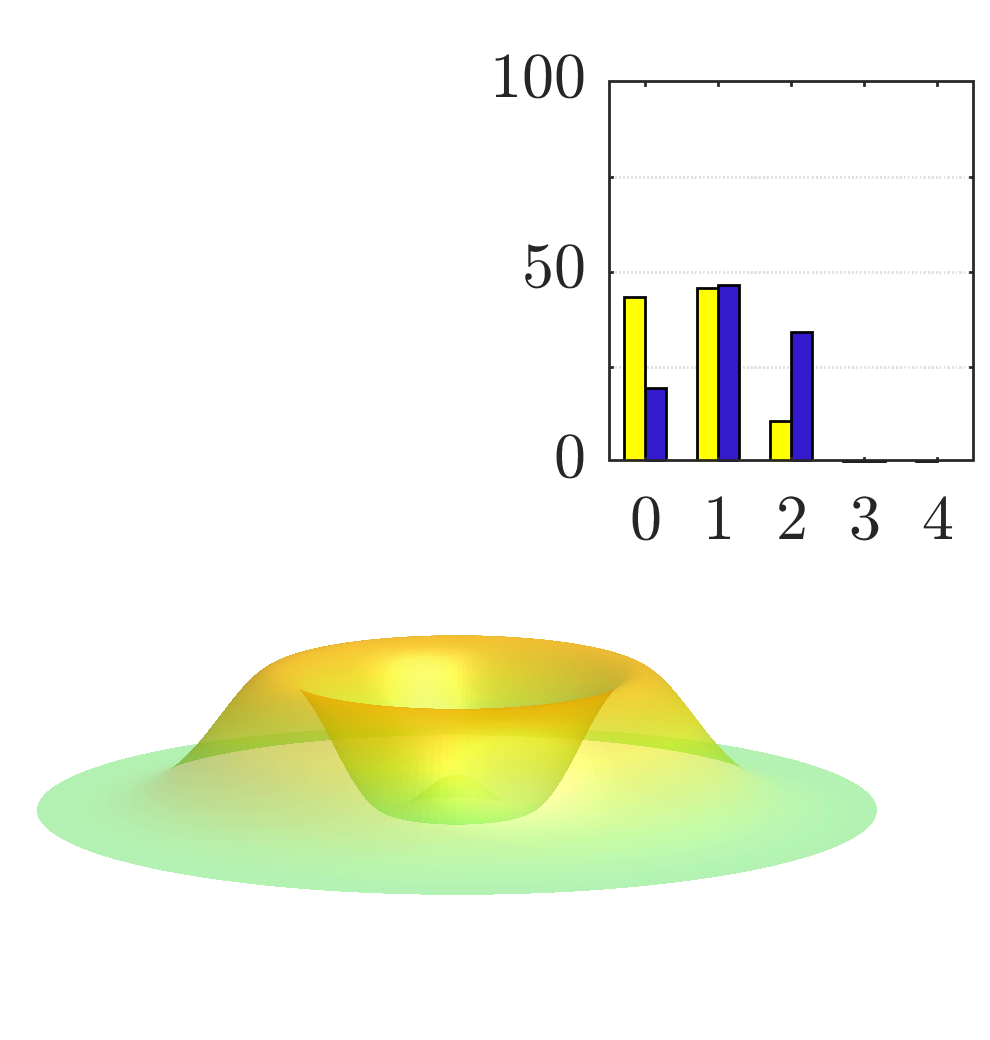}
    		\caption{40 round trips \\ (420 ns)}
    	\end{subfigure}

		\caption{Wigner function of the single-photon (top) and two-photon state (bottom) after different storage round trips: from 0~(a) to 40~(e) round trips. The Wigner functions are corrected from a \SI{77}{\percent} detection efficiency. Inset: diagonal elements of the density matrices without (yellow) and with (purple) correction of the detection efficiency.}
		\label{fig:WF}
	\end{figure*}

A scheme of the experimental setup is shown in figure~\ref{scheme}. Pulses from a Ti:sapphire laser with a central wavelength of \SI{850}{\nano\meter} (temporal length: \SI{2.6}{\pico\second} - repetition rate: \SI{76}{\mega\hertz}) are frequency doubled using a $\sim \SI{3.9}{\meter}$-long bow-tie second-harmonic-generation cavity with a type I \bibo crystal~\cite{KanseriBouillardTualle-Brouri2016,Bouillard2018}. The cavity, of finesse~60, allows to enhance the non-linear effect, leading to a conversion efficiency of \SI{70}{\percent}. The output pulses at \SI{425}{\nano\meter} are sent to another $\sim\SI{3.9}{\meter}$-long bow-tie cavity of finesse $\simeq 80$, namely the optical parametric amplification (OPA) cavity. The intra-cavity peak power reaches~\SI{40}{\kilo\watt}. A second type I \bibo crystal placed inside the OPA cavity allows to produce pairs of photons by non-colinear spontaneous parametric down conversion (SPDC) \cite{Bouillard2018}.

As the phase matching of SPDC is broadband, the heralding part of the photon pairs, in the signal beam, is first spectrally and spatially filtered by the use of a single mode fiber and a grating combined with a slit. The filtered photons are probabilistically separated by a beam-splitter followed by two single-photon counting modules (SPCM, Perkin Elmer SPCM-AQR-13) in order to detect the generation of single and two-photon Fock states.

To account for the delay of the electronics ($\sim\SI{200}{\nano\second}$), the heralded state in the idler beam is sent through an optical delay-line consisting in an open \SI{4.8}{\meter} long cavity made of three mirrors \cite{Bouillard2018}. It allows to tune the delay from 32 to \SI{288}{\nano\second} by steps of \SI{32}{\nano\second} with optical losses lower than \SI{1}{\percent}. The H-polarized Fock state is then transmitted by the PBS of the quantum memory and enters into the cavity. A \SI{14}{\nano\second} high-voltage pulse is applied to the Pockels cell in order to turn the polarization of the light pulse to vertical, and the Fock state is then stored inside the cavity. After a given number of round trips, the quantum state is extracted out of the cavity by rotating again its polarization. The extracted state is sent to a homodyne detection in order to be characterized, the overall efficiency $\eta$ of the homodyne detection being of $\eta_{\mathrm{HD}}=\eta_{\mathrm{PD}}\eta_\mathrm{C}=77\pm 3\%$ (where $\eta_{\mathrm{PD}}=94\pm2\%$ is the quantum efficiency of the photodiodes, and $\eta_{\mathrm{C}}$=C$^2$=81$\pm$1\% is the mode-matching efficiency, with C the contrast measured between the idler beam and the local oscillator \cite{leonhardt1997measuring,LvovskyHansenAicheleEtAl2001}).

To avoid the saturation of our homodyne detection (bandwidth of a few~\si{\mega\hertz}) we reduce the repetition rate of the local oscillator. To do so, another Pockels cell (PC) combined with a polarizer is used and work synchronously with the quantum memory. Thus a pulse from the local oscillator is sent to the homodyne detection only when a quantum state is extracted from the quantum memory. As the production rate of the Fock states is low compared to the bandwidth of the detector, the analysis and production rates are almost identical \cite{Bouillard2018}.

The equation \ref{eq:prob} shows that the probability to synchronize two quantum states is defined by the number of round trips allowed in the cavity and by the production rate of these quantum states. The losses per round-trip of the cavity are estimated to be \SI{0.6}{\percent}, leading to a theoretical lifetime of a single-photon of $\tau=\SI{2.2}{\micro\second}$ (defined as $\mathcal{F}=\mathcal{F}_0 e^{-t/ \tau}$, where $\mathcal{F}_0$ denotes the initial fidelity of the state~\footnote{$\mathcal{F}=\mathrm{tr}\left(\rho\ket{\psi}\bra{\psi}\right)$ where $\ket{\psi}$ is the reference state and $\rho$ the experimental density matrix~\cite{Jozsa1994}.}). The initial fidelity of the Fock states (without storage) is another key parameter of the setup: the quality of the single-photon (two-photon) states reaches $91\pm\SI{4}{\percent}$ ($85 \pm\SI{6}{\percent}$) after correction from the detection efficiency $\eta$, along with a production rate of 150-\SI{250}{\kilo\hertz} ($\sim\SI{200}{\hertz}$)\cite{Bouillard2018}.

\section{Results and Discussion}

To characterize the quantum memory, homodyne tomographies of the single-photon and two-photon states have been realized for different numbers of round trips inside the cavity. For a given number of round trips, \SI{50000}{} and \SI{10000}{} data points have been acquired for respectively the single-photon and two-photon measurements. As the Fock states are phase invariant, the phase of the homodyne detection was not measured. Once acquired, the density matrix and the Wigner function of the stored states, represented on the figure \ref{fig:WF}, are reconstructed using a maximum likelihood technique \cite{Lvovsky2004} with correction of the detection efficiency $\eta$. The diagonal elements of the density matrices with and without correction of the efficiency are also shown in the insets of the figure \ref{fig:WF}. As they increase with the number of round trips, the effect of losses in the quantum memory reduces the fidelity of the stored states leading to a decrease of the negativity of the Wigner function. Both states tend towards a bi-dimensional Gaussian shape representing the Wigner function of the vacuum.

	\begin{figure*}[t]

		\centering		
		\begin{subfigure}{0.48\linewidth}
			\includegraphics[width=1\linewidth]{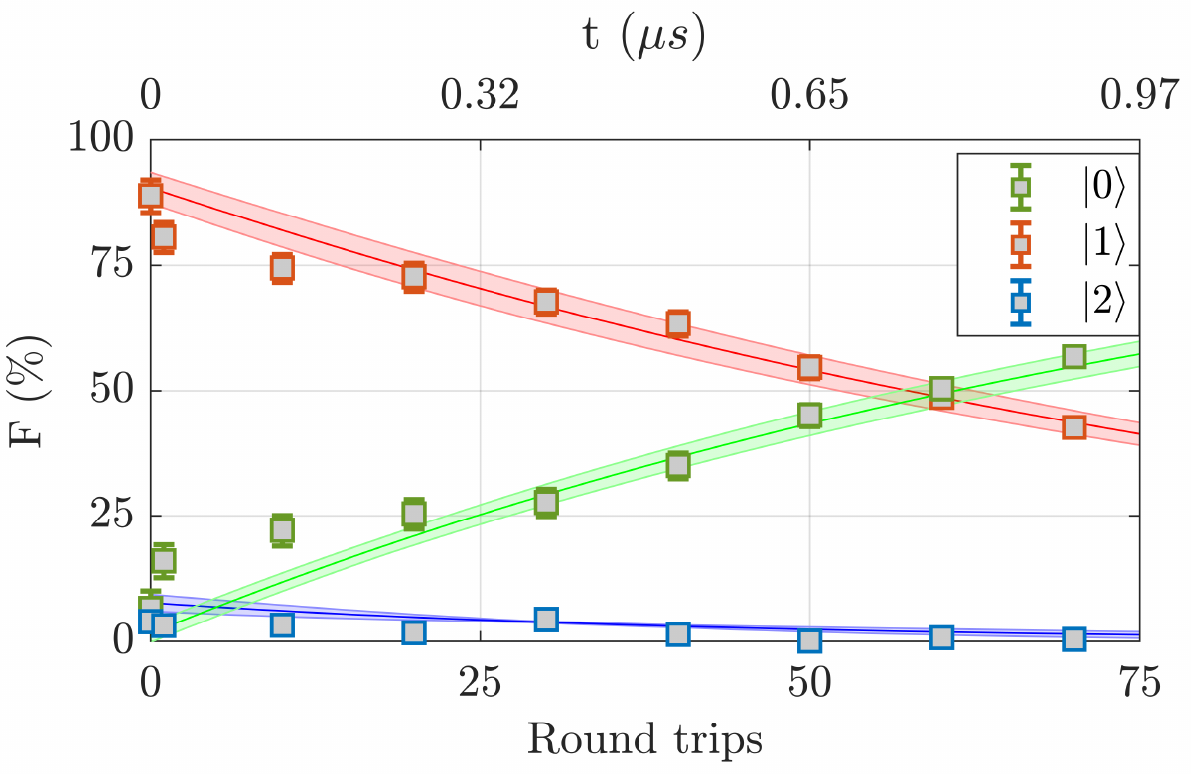}
    		\caption{}
    		\label{Delta fit sp}
    	\end{subfigure}
        \qquad
		\begin{subfigure}{0.48\linewidth}
			\centering
			\includegraphics[width=1\linewidth]{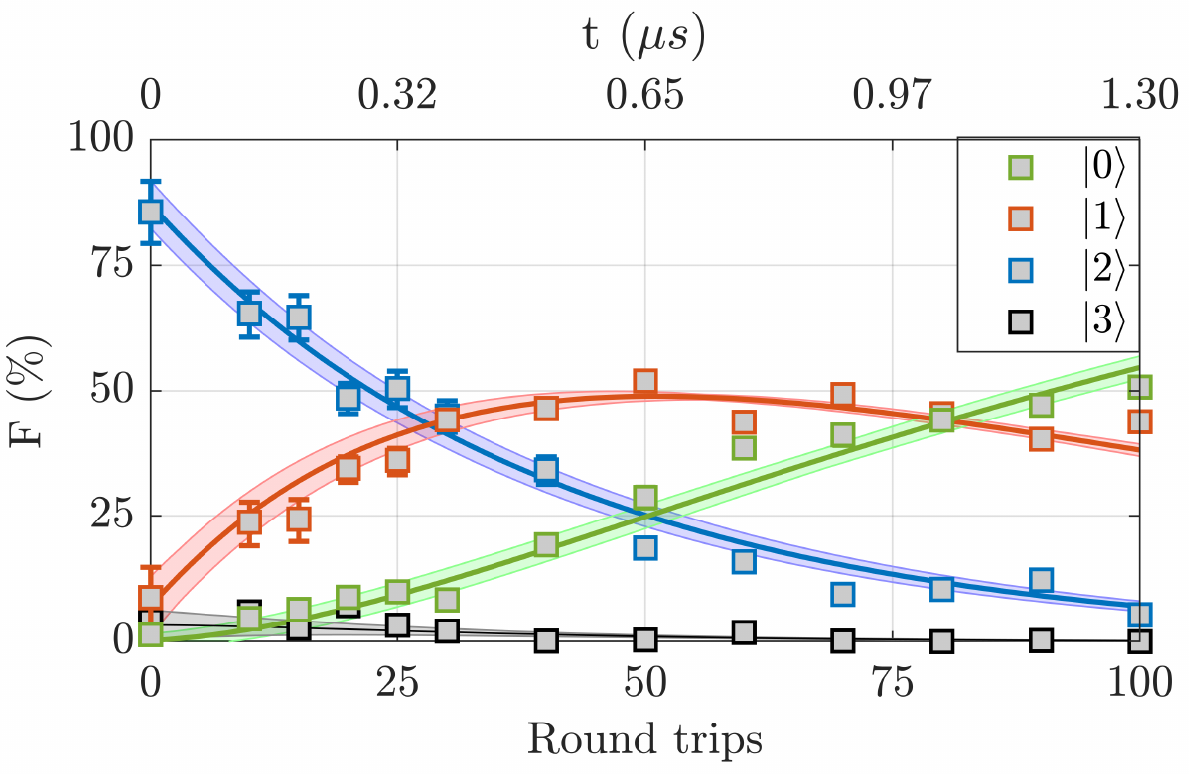}
   			 \caption{}
			 \label{Delta fit 2ph}
        \end{subfigure}

		\caption{Evolution of the diagonal elements of the density matrix of the stored state for the single-photon (a) and two-photon Fock state (b), as a function of the number of round trips (bottom axis) and of the storage time (top axis). The coloured bands around the fits account for the uncertainty on the detection efficiency.}
		\label{fig:el_diag}
	\end{figure*}

\begin{figure}
	\includegraphics[width=1\linewidth]{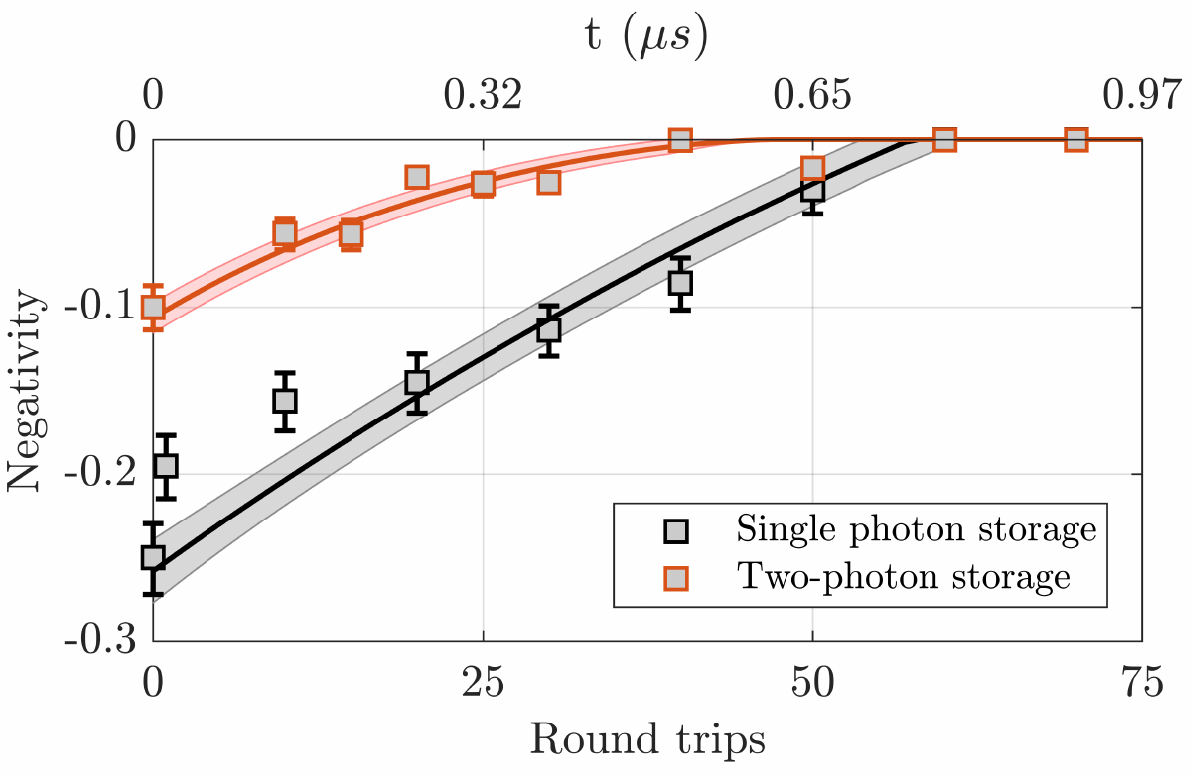}
%
%
    \caption{Evolution of the negativity as a function of the number of round trips (bottom axis) and of the storage time (top axis). The coloured bands around the fits account for the uncertainty on the detection efficiency.}
    \label{negat_fit}
\end{figure}

The values of the diagonal elements of the density matrix (corrected from the detection efficiency), which are the fidelities $\mathcal{F}_{\ket{n}}=\bra{n}\rho\ket{n}$ of the stored state $\rho$ with the Fock state $\ket{n}$, are represented on the figure~\ref{fig:el_diag} for the single-photon Fock states (figure \ref{Delta fit sp} : red) and the two-photon Fock state (figure \ref{Delta fit 2ph} : blue) as a function of the number of round trips of storage. The lifetimes are estimated to be of \SI{1.44}{\micro\second} and \SI{0.51}{\micro\second} for the single-photon and the two-photon states. Such a difference is explained by the quadratic effect of the losses on the two-photon states.
As it reveals the non-classicality of quantum states, the negativity of the Wigner function is a more relevant criteria than the lifetime for the characterization of memories in the continuous variable domain. The negativity of the reconstructed Wigner functions is directly plotted on the figure~\ref{negat_fit}. Starting at -0.25$\pm0.02$ for the single-photon and -0.10$\pm0.02$ for the two-photon state, the negativity is maintained over 50 round trips for both states.
In both figures~\ref{fig:el_diag} and~\ref{negat_fit}, error bars are deduced from the uncertainty on the detection efficiency $\eta$, which is the major error source in this experiment.

In order to evaluate the losses of the cavity, the evolution of the density matrix were fitted using the equation developed in~\citep{KissHerzogLeonhardt1995}:
    \begin{multline}
            \bra{n}\hat{\rho}_{N}\ket{n}=(1-p)^{Nn}\sum_{i=0}^{\infty} \binom{n+i}{n}(1-(1-p)^N)^i\\
            		\bra{n+i}\hat{\rho}_{0}\ket{n+i}
            \label{eq:kiss_qm_simp}
    \end{multline}

Where $p$ is the losses per turn of the cavity and $\rho_N$ is the density matrix of the states after N round trips of storage. The fits of the data in figures \ref{fig:el_diag} show a good agreement with experimental data points. The losses per round trip can be extracted from the fits, and are estimated to be of \SI{1.0}{\percent} for the single-photon and \SI{1.3}{\percent} for the two-photon state experiment, exhibiting a slight decrease for the two-photon states. Such a difference could be explained by a longer acquisition time for the two-photon states, making the results more sensitive to experimental fluctuations. The losses per round trip do not only include the optical losses of the cavity, but also the evolution for the spatial mode of the quantum state inside the cavity as well as temporal mismatch between the cavity length and the repetition rate of the laser, leading to temporal and spatial mismatches between the local oscillator and the quantum state. The optical losses per round trip are estimated to be of \SI{0.6}{\percent}, which is close to the experimental values, showing a good control of the spatio-temporal mode of the stored quantum states.
The fits allows the determination of the negativities for each number of round trips of storage. For the single-photons, this negativity starts at $-0.26\pm\SI{0.03}{}$, and the Wigner function exhibit negativity after 57$\pm$4 round trips of storage, corresponding to a $0.75\pm\SI{0.05}{\micro\second}$ storage time with preservation of this non-classical feature. For the two-photon states the Wigner functions remain negative until 46$\pm$6~round trips, leading to a storage time of $0.60\pm\SI{0.08}{\micro\second}$.

With the lifetimes of the cavity, we can now estimate the probability of synchronizing two single-photon Fock states (equation~\ref{eq:prob}). This value is of \SI{14}{\percent}, leading to a synchronization rate of two single-photon states of approximately~\SI{28}{\kilo\hertz}, corresponding to a rate enhancement of the synchronization by a factor of about $53$, which is to our knowledge the highest value reported for such kind of memory. Compared to our previous experiment with two independent single-photon sources \cite{etesse2015experimental}, this would correspond to an increase by more than 4 order of magnitude. As it can be seen from equation~\ref{eq:prob}, the production rate is a key parameter for synchronizing two quantum states. In order to increase the heralding rate, one can increase the detection efficiency of the heralding part of the experiment (\SI{6}{\percent}), by using high-efficiency single-photon detectors \cite{JeannicVermaCavaillesEtAl2016,FukudaFujiiNumataEtAl2011} or by reducing the losses of the spectral filtering (currently of $\sim \SI{15}{\percent}$), increasing significantly the success rate of the synchronization. As an example, an increase of the detection efficiency by a factor 3 would give a success rate of~$\sim\SI{200}{\kilo\hertz}$ for a success probability of $\sim 36\%$.

In conclusion, we have studied the storage of single and two-photon Fock states in an all-optical cavity, reporting what is, to our knowledge, the first storage of a non-Gaussian state with more than one photon. The stored quantum states show negative Wigner functions after 57$\pm$4 and 46$\pm$6~round trips for single and two-photon states, respectively, which corresponds to a storage time of respectively $0.75\pm\SI{0.05}{\micro\second}$ and $0.60\pm\SI{0.08}{\micro\second}$. Such cavities pave the way towards the synchronization of optical quantum states, which is a cornerstone in the direction of quantum information protocols.






%

\end{document}